\documentclass[aps,groupedaddress,prd,twocolumn,eqsecnum,epsbox]{revtex4}

\usepackage{epsfig}

\usepackage{amsmath,amssymb}
\usepackage{color}

\begin{document}

\title{Holographic bound in covariant loop quantum gravity}

\author{Takashi Tamaki}
\affiliation{Department of Physics, General Education, College of Engineering, 
Nihon University, Tokusada, Tamura, 
Koriyama, Fukushima 963-8642, Japan}
\email{tamaki@ge.ce.nihon-u.ac.jp}


\begin{abstract}
We investigate puncture statistics based on the covariant area spectrum in loop quantum gravity. 
First, we consider Maxwell-Boltzmann statistics with a Gibbs factor for punctures. 
We establish formulae which relate physical quantities such as horizon area to the parameter 
characterizing holographic degrees of freedom. 
We also perform numerical calculations and obtain consistency with these formulae. 
These results tell us that the holographic bound is satisfied in the large area limit 
and correction term of the entropy-area law 
can be proportional to the logarithm of the horizon area. 

Second, we also consider Bose-Einstein statistics and show that the above formulae are also 
useful in this case. By applying the formulae, we can understand intrinsic features of 
Bose-Einstein condensate which corresponds 
to the case when the horizon area almost consists of punctures in the ground state. 
When this phenomena occurs, 
the area is approximately constant against the parameter characterizing the temperature. 
When this phenomena is broken, the area shows rapid increase which suggests the phase transition 
from quantum to classical area. 
\end{abstract}

\date{\today}

\maketitle

\section{Introduction}
Canonical quantization of general relativity has a long history beginning in 
the 1960s~\cite{Wheeler}. Basically, metric and its conjugate momentum had been  
used as canonical variables in these early days. 
In this case, the Hamiltonian constraint is nonpolynomial about these variables. 
So, it is almost impossible to solve its quantized counterpart. 
In \cite{complex}, it was shown that if we use the complex Ashtekar connection and its 
conjugate, the Hamiltonian constraint can be written as polynomial about 
these variables. Surprisingly, it was found that Wilson loop for this connection 
is a solution of quantized 
Hamiltonian constraint~\cite{Jacobson}. Using Wilson loop, spin network, basic 
ingredients of the loop quantum gravity(LQG), has been constructed~\cite{Smolin}. 
Discrete area spectrum is one of the main predictions in LQG~\cite{Rovelli, Ash1}.

However, it has been recognized that the reality conditions for the physical 
quantities to be real are difficult to be solved. Then, the SU(2) real connection  
has been introduced where imaginary number $i$ in the complex Ashtekar connection 
was replaced by the real parameter $\gamma$ 
called the Barbero-Immirzi(BI) parameter~\cite{Immirzi}. 
Although the Hamiltonian constraint becomes nonpolynomial, 
this complication can be relieved if we rewrite it using the technique developed  
in~\cite{Thiemann's method}. If we apply it in the symmetry-reduced model, 
we can discuss singularity avoidance 
which has been paid much attention~\cite{LQC}. 

The microscopic origin of black hole entropy in LQG 
had also been discussed in \cite{Rovelli-entropy}, where the number of 
degrees of freedom of the edge configuration for a fixed SU(2) area spectrum was counted. 
Then, Ashtekar,  Baez, Corichi, and Krasnov refined this idea based on 
the isolated horizon framework (so-called ABCK framework 
where the isolated horizon itself was described by 
U(1) connection~\cite{ABCK,isolated}) 
and determined $\gamma$ to satisfy the Bekenstein-Hawking entropy-area law 
$S=A/(4G)$ where $S$, $A$ and $G$ are black hole entropy, horizon area 
and the gravitational constant, respectively. Here, ambiguity of $\gamma$ turned out to 
be the merit of using the real connection. 
Including the correction of error in original counting \cite{Domagala}, or ambiguity in 
counting \cite{Alekseev,Mitra,Tamaki}, relation with the quasinormal 
mode~\cite{Schiappa,Dreyer,Hod}, various aspects have been discussed 
related to the ABCK framework~\cite{projection,Sahlmann,Ansari}. 

The situation slightly changed when it was found that 
the isolated horizon can be written using the SU(2) connection~\cite{Engle}. 
This means that the horizon Hilbert space can be described by the SU(2) Chern-Simons 
state. Its dimension is written by the spin freedom $j$ and the level of the 
Chern-Simons state $k$. Using a suitable analytic continuation of these variables to 
complex variables, it was obtained that the complex Ashtekar connection is desirable 
to reproduce $S=A/(4G)$~\cite{BTZ,analytic}. Furthermore, when we introduce the 
geometric temperature by demanding the horizon state be a Kubo-Martin-Schwinger 
state, we can also arrive at the complex connection~\cite{temperature}. 

Is there an essential reason why the complex Ashtekar connection is preferable?
One of the reasons would be that the covariance is satisfied in this connection 
while it is violated in the real connection \cite{Samuel}. 
This should be taken seriously, and we should pay attention how to choose 
the Lorentz covariant connection which has been investigated in \cite{Choice}. 
The connection obtained in \cite{Choice} is called shifted connection which 
includes the BI parameter. Surprisingly, the Hamiltonian constraint can be written 
as a polynomial equation in this case again. 
Using a shifted connection, covariant LQG has been formulated, and covariant 
area spectrum has been obtained~\cite{areaCLQG,HilbertCLQG}. 
Making the consistent relation between covariant LQG and the spin foam models 
became the important realm recently~\cite{reviewCLQG}. 
We should also notice that the covariant area spectrum does not include 
the BI parameter although the shifted connection itself does. 
Then, it is natural to ask whether or not we can obtain consistency with 
the entropy-area law 
if we consider counting microscopic freedom of black holes in covariant LQG. 
In \cite{countingCLQG},  by assuming the horizon area consists of the minimum area 
eigenvalue, it was argued that the answer is in the affirmative. 

Here, we consider the generality of the holographic bound and argue the 
correction term of the entropy-area law discussed in \cite{first,gas}. 
These are motivated by the quasilocal first law of black hole thermodynamics where 
the quasilocal energy is defined using the horizon area~\cite{quasilocal}. 
Then, regarding the puncture, which is an intersection of the edge at the horizon, 
as a particle, we can argue its statistical mechanics. One of the important points 
in \cite{first,gas} is that if we assume the degeneracy of matter fields 
close to the horizon as $\exp (\lambda A/G)$ where $\lambda$ is a 
dimensionless constant, $\lambda$ must approach $1/4$ in the large area limit 
when punctures are indistinguishable. The correction term of the entropy-area law 
is basically proportional to $\sqrt{A}$ unless we assume the special form for 
the fugacity. Then, our concerns are 
whether these properties hold or not in the covariant area spectrum. 
The answer is in the affirmative for the holographic bound while the 
correction term depends on the ambiguity of the covariant area spectrum 
as we discuss later. 

This paper is organized as follows. 
In section II, we introduce tools necessary for constructing puncture 
statistics following \cite{first,gas}.  In Sec. III, 
we consider the case when Maxwell-Boltzmann statistics with 
a Gibbs factor for punctures is assumed. 
We establish formulae which relate physical quantities such as horizon area to the parameter 
characterizing holographic degrees of freedom. We also perform numerical calculations and 
obtain consistency with these formulae. These results show that the holographic bound is saturated in the 
large area limit and that the correction term of the entropy-area law can 
be proportional to $\ln A$. In section IV, we consider the 
case when Bose-Einstein statistics is assumed and argue that the above formulae are also 
useful in this case. By applying the formulae, we can understand the intrinsic features of 
Bose-Einstein condensate which corresponds 
to the case when black holes almost consist of punctures in the ground state. 
We show that when this phenomena occurs, 
the area is approximately constant against the parameter characterizing the temperature. 
When this phenomena is broken, the area shows rapid increase which suggests the phase transition 
from quantum to classical area spectrum. 
Concluding remarks follow in section V.

\section{Preparation for puncture statistics}
Following \cite{first,gas}, we introduce several notions necessary for arguing puncture statistics. 
First, we mention the quasilocal law of black hole thermodynamics which holds for 
the stationary observer at proper distance $l$ from the horizon~\cite{quasilocal}, 
\begin{eqnarray}
E=\frac{A}{8\pi l}, 
\label{quasilocal-1st}
\end{eqnarray}
where $E$ is quasilocal horizon energy of black hole. 
We rewrite (\ref{quasilocal-1st}) using the inverse Unruh temperature 
$\beta_{\rm U}:=\frac{2\pi l}{G}$ as 
\begin{eqnarray}
\beta_{\rm U}E=\frac{A}{4G}\ . 
\label{quasilocal-Unruh}
\end{eqnarray}

Then, we can discuss the energy spectrum of the puncture by combining (\ref{quasilocal-Unruh}) with 
the area spectrum. In \cite{first,gas}, the SU(2) area spectrum written as 
\begin{eqnarray}
A=8\pi \gamma G\sum_{i}\sqrt{j_{i}(j_{i}+1)}, 
\label{SU(2)}
\end{eqnarray}
has been used. Here, $j_{i}$ is a half-integer associated with the puncture $i$. 
Here, we use the covariant area spectrum written as 
\begin{eqnarray}
A=8\pi G\sum_{i}\sqrt{j_{i}(j_{i}+1)-n_{i}^{2}+\rho_{i}^{2}+1}\ , 
\label{general}
\end{eqnarray}
where $n_{i}$ is a half-integer with $j_{i}\geq n_{i}$ and $\rho_{i}$ is a 
real number~\cite{areaCLQG}. Notice that there is no ambiguity related 
to $\gamma$. In \cite{HilbertCLQG}, 
it has been shown that it is enough for counting the degrees of freedom 
to consider the simple representation $n_{i}=0$, which we assume here. 

The important point is how to determine $\rho_{i}$. 
The relation (\ref{quasilocal-1st}) and the spectrum (\ref{general}) show 
that the puncture $i$ has quasilocal energy 
\begin{eqnarray}
E_{i}=\frac{G}{l}\sqrt{j_{i}(j_{i}+1)+\rho_{i}^{2}+1}\ . 
\label{each-puncture-energy}
\end{eqnarray}
Thus, the simplest possibility is to choose $\rho_{i}=0$, which we include 
considering below. The next simplest possibility would be to regard $\rho_{i}$ as a 
dependent variable of $j_{i}$. In this case, $j_{i}=1/2$ does not necessarily 
correspond to the ground state, which is important when we discuss Bose-Einstein 
condensate as shown in \cite{gas}. 
Although it is an interesting possibility, 
it is reasonable to assume that $E_{i}$ is monotonic with $j_{i}$ as a first 
extension of the previous case in \cite{first,gas}. 
Here, we choose $\rho_{i}^{2}$ as  
\begin{eqnarray}
\rho_{i}^{2}=0,\ j_{i}^{2m}\ (m>1),\  e^{2j_{i}}, 
\label{function-form0}
\end{eqnarray}
which correspond to the cases, 
\begin{eqnarray}
\frac{l}{G}E_{i}\to j_{i},\ j_{i}^{m},\ e^{j_{i}},
\label{energy-asymptotic0}
\end{eqnarray}
in the limit $j_{i}\to\infty$, respectively. The reason why we choose a monomial or 
an exponential as 
(\ref{function-form0}) is supposed by the observation that only the qualitative behavior 
in the limit $j_{i}\to\infty$ determines the holographic property and the 
correction term of the entropy-area law in \cite{first,gas}. 

Let us consider puncture statistics. 
In general, we do not require that 
the inverse temperature is equal to $\beta_{\rm U}$. 
We write the inverse temperature $\beta$ using $\beta_{\rm U}$ as 
\begin{eqnarray}
\beta =\beta_{\rm U}(1+\delta_{\beta})\ ,
\label{Unruh}
\end{eqnarray}
where $\delta_{\beta}$ is a parameter. We only demand that $\delta_{\beta}$
vanishes in the semiclassical limit $A\to \infty$ to 
satisfy the relation (\ref{quasilocal-Unruh}). 

We define $n_{j}$ as the number of punctures carrying spin $j$ and 
$N$ as the total number of punctures. So, we have 
\begin{eqnarray}
N=\sum_{j}n_{j}\ .
\label{total}
\end{eqnarray}
We also define $D(\{n_{j}\})$ as the number of holographic degrees of freedom 
for a given configuration $\{n_{j}\}$. 
Here, we assume 
\begin{eqnarray}
D(\{n_{j}\})=\exp \left( (1-\delta_{h})\bar{A}\right)\ ,
\label{degeneracy}
\end{eqnarray}
where $\bar{A}:=\frac{A}{4G}$ and $\delta_{h}$ is a free parameter. We suppose 
that the freedom $D(\{n_{j}\})$ comes from 
the matter fields close to the horizon motivated by the entanglement 
entropy hypothesis~\cite{Bombelli}. 

\section{Maxwell-Boltzmann statistics}

We include the Gibbs factor $N!$ in 
the Maxwell-Boltzmann statistics. The case without the Gibbs factor is discussed 
later in this section. Then the canonical partition function 
$Q(N,\beta)$ is given by 
\begin{eqnarray}
Q(N,\beta )=\frac{1}{N!}\sum_{n_{j}}D(\{n_{j}\})\frac{N!}{\prod_{j} n_{j}!}
\prod_{j}e^{-\beta n_{j}E_{j}}\ .
\label{canonical-partition}
\end{eqnarray}
Here, we abbreviate the puncture index $i$ and write the spin index $j$ in the quasilocal 
energy as 
\begin{eqnarray}
E_{j}=\frac{G}{l}\sqrt{j(j+1)+\rho^{2}+1}\ .
\label{E-spin}
\end{eqnarray}
Using (\ref{degeneracy}), we can express the partition function as 
\begin{eqnarray}
Q=\frac{q^{N}}{N!}\ ,
\label{canonical-partition2}
\end{eqnarray}
where 
\begin{eqnarray}
q=\sum_{j=1/2}^{\infty}\exp (-2\pi\delta\sqrt{j(j+1)+\rho^{2}+1})\ .
\label{correction-term}
\end{eqnarray}
Here, we defined 
\begin{eqnarray}
\delta :=\delta_{\beta}+\delta_{h}\ .
\label{delta}
\end{eqnarray}

We introduce the fugacity $z=\exp (\beta\mu )$ where $\mu$ is a chemical 
potential. In this case, we can express 
the grand canonical partition function by 
\begin{eqnarray}
Z_{\rm MB}=\sum_{N}\frac{(zq)^{N}}{N!}=\exp (zq)\ .
\label{grand-canonical}
\end{eqnarray}
The total number $N$ and the mean energy $E$ are 
\begin{eqnarray}
&&N=z\frac{\partial}{\partial z}(\ln Z_{\rm MB})=zq\ , \label{N} \\
&&E=-\partial_{\beta}(\ln Z_{\rm MB})+N\mu =-z\partial_{\beta}q\ .
\label{mean-energy}
\end{eqnarray}
We can express the entropy as 
\begin{eqnarray}
&&S=\beta E-N\ln z+\ln Z_{\rm MB} \nonumber  \\
&&=\bar{A}(1+\delta_{\beta})+N(1-\ln z)\ .
\label{mean-entropy0}
\end{eqnarray}
As we said above, we required $\delta_{\beta}\to 0$ in the limit $A\to\infty$. 
So, if $z=e$, the correction term of the entropy-area law proportional 
to $N$ disappears as pointed out in \cite{gas}. 

Since one of the purposes using 
the covariant area spectrum is to investigate the correction term, 
we consider the case $z\neq e$. 
In other treatments, it is often argued that the correction term 
proportional to $\ln \bar{A}$ appears~\cite{Carlip,Das,Sen}. 
From (\ref{N}) and (\ref{mean-energy}), we have 
\begin{eqnarray}
\bar{A}:N(1-\ln z)=-\beta\partial_{\beta}q:q(1-\ln z)\ .
\label{ratio}
\end{eqnarray}
Thus, in discussing the ratio between $\bar{A}$ and the correction term, 
it is enough if we investigate the ratio between $\partial_{\beta}q$ and $q$. 
Since $z$ plays a minor role for this reason, we set $z=1$ below, for simplicity.

How can we estimate the relation between $q$ and $A$ ? 
We should first notice that convergence of the sequence (\ref{correction-term}) 
highly depends on $\delta$. So, our strategy is to analyze the dependencies of 
$q$ and $A$ as a function of $\delta$ for obtaining 
the relation between $q$ and $A$. 

Since it would be difficult in calculating (\ref{correction-term}) exactly, 
we suppose using numerical calculation. In this case, 
it is important to know $j_{\rm max}$ 
we should sum up, which is a key to understand above property. 
In concrete, we assume that we need to sum up from $j=1/2$ to $j_{\rm max}$ in 
obtaining the value $q_{\rm fix}$ for enough precision toward the true value $q$, 
e.g., relative error $|q-q_{\rm fix}|/q<10^{-20}$. To accomplish the above task, 
we need to estimate the dependence of $j_{\rm max}$ on $\delta$ as a first step, 
which is also a difficult task, in general. However, 
we can expect that $j_{\rm max}\to\infty$ in the limit $\delta\to 0$, 
and we can estimate (\ref{correction-term}) using 
the asymptotic form in the limit $j\to\infty$. 
For this reason, we assume $|\delta|\ll 1$. 

Let us consider the cases (\ref{energy-asymptotic0}). 
If we have $\bar{E}_{j}:=\frac{l}{G}E_{j}\to j^{n}$ $(n=1,m)$ in $j\to\infty$, 
we can write as 
\begin{eqnarray}
\hspace{-5mm}q_{\rm fix}\simeq\sum_{j=1/2}^{j_{\rm max}}\exp (-2\pi\delta j^{n})
=\sum_{j=1/2}^{j_{\rm max}}\exp \left[-2\pi (\epsilon j)^{n}\right]\ ,
\label{correction-evaluate}
\end{eqnarray}
where $\delta =:\epsilon^{n}$. If we define $x:=\epsilon j$ and 
$f(x):=\exp \left(-2\pi x^{n}\right)$, we can rewrite as 
\begin{eqnarray}
q_{\rm fix}\simeq\sum_{x=x_{1}}^{x_{2}}f(x)\ ,
\label{correction-key}
\end{eqnarray}
where $x_{1}=\epsilon /2$ and $x_{2}=\epsilon j_{\rm max}$. 

Using these notations, we comment on following important properties.  
\begin{itemize}
\item If we reduce $\epsilon_{\rm old}\to \epsilon_{\rm new}=\epsilon_{\rm old}/10$, 
\begin{list}{}{}
\item{(i)} we should change $j_{\rm max,old}\to j_{\rm max,new}=10j_{\rm max,old}$ in 
preserving the same precision. 
\item{(ii)} we obtain $q_{\rm fix,old}\to q_{\rm fix,new}=10q_{\rm fix,old}$ 
approximately. 
\end{list}
\end{itemize}
To understand these properties, we should first notice that 
interval $\Delta x=\frac{\epsilon}{2}$ in the sum (\ref{correction-key}) becomes $1/10$ 
while $x_{2}$ does not change by (i). This means 
that there are $2j_{\rm max,old}$ terms we should sum up in the former case 
while $20j_{\rm max,old}$ terms in the latter case in (\ref{correction-key}). 
Thus, we obtain $q_{\rm fix,old}\to q_{\rm fix,new}=10q_{\rm fix,old}$ 
approximately. Since $|q-q_{\rm fix}|=\sum_{x_{2}}^{\infty}f(x)$, 
we also have $|q_{\rm old}-q_{\rm fix,old}|\to 10|q_{\rm old}-q_{\rm fix,old}|$ 
approximately. Therefore, we have same relative error and the precision is preserved. 

For this approximation to be valid, following conditions should hold. 

\vspace{5mm}
{\bf Conditions}

\begin{itemize}
\item Changing $x_{1,\rm old}\to x_{1,\rm new}$ is negligible. 
\item $f(x)$ does not have the property, 

$|f(x+\Delta x)/f(x)|\ll 1$, or $\gg 1$. 
\end{itemize}

The former assumption is implicitly used when we use the asymptotic form in the 
limit $j\to\infty$. The latter assumption holds when $\delta$ is small enough in 
the above case. 

From these consideration,  we obtain  
\begin{eqnarray}
q\propto\epsilon^{-1}=\delta^{-1/n}\ .
\label{delta-q-relation}
\end{eqnarray}
Since $A\propto -\partial_{\beta}q \propto -\partial_{\delta}q$, we also have 
\begin{eqnarray}
A\propto\delta^{-(n+1)/n}\ .
\label{delta-A-relation}
\end{eqnarray}
We mention that our results (\ref{delta-q-relation}) and 
(\ref{delta-A-relation}) are consistent with those in \cite{gas}
where (\ref{SU(2)}) was used which corresponds to the case $n=1$. 

Next, we consider the case $\bar{E}_{j}\to e^{j}$ in $j\to\infty$. 
In this case, we can write as 
\begin{eqnarray}
q_{\rm fix}\simeq\sum_{j=1/2}^{j_{\rm max}}\exp (-2\pi\delta e^{j})\ .
\label{correction-evaluate2}
\end{eqnarray}
As in the previous case, 
if we want to obtain $q_{\rm fix}\to Bq_{\rm fix}$ $(B\gg 10)$, 
we need to change the number of terms we should sum up from $2j_{\rm max}$ to 
$2Bj_{\rm max}$ $(B\gg 10)$ for preserving the precision. 
This means that $\delta$ should change to satisfy 
\begin{eqnarray}
\delta_{\rm old} e^{j_{\rm max}}= \delta_{\rm new} e^{Bj_{\rm max}}\ .
\end{eqnarray}
So we have $\delta_{\rm new}=\delta_{\rm old} e^{(-B+1)j_{\rm max}}
\simeq \delta_{\rm old} e^{-Bj_{\rm max}}$. 
This means $B\simeq -\frac{1}{j_{\rm max}}\ln 
\left(\frac{\delta_{\rm new}}{\delta_{\rm old}}\right)$. 
As a result, we have 
\begin{eqnarray}
q\propto -\frac{1}{j_{\rm max}}\ln 
\left(\frac{\delta}{C}\right)\ ,
\label{delta-q-relation2}
\end{eqnarray}
where $C$ is a constant. So we have 
\begin{eqnarray}
A\propto -\frac{1}{j_{\rm max}\delta}\ .
\label{delta-A-relation2}
\end{eqnarray}
The formulae (\ref{delta-q-relation}), (\ref{delta-A-relation}), (\ref{delta-q-relation2}), 
and (\ref{delta-A-relation2}) play quite important roles in this paper. 

\begin{figure}[b]
\psfig{file=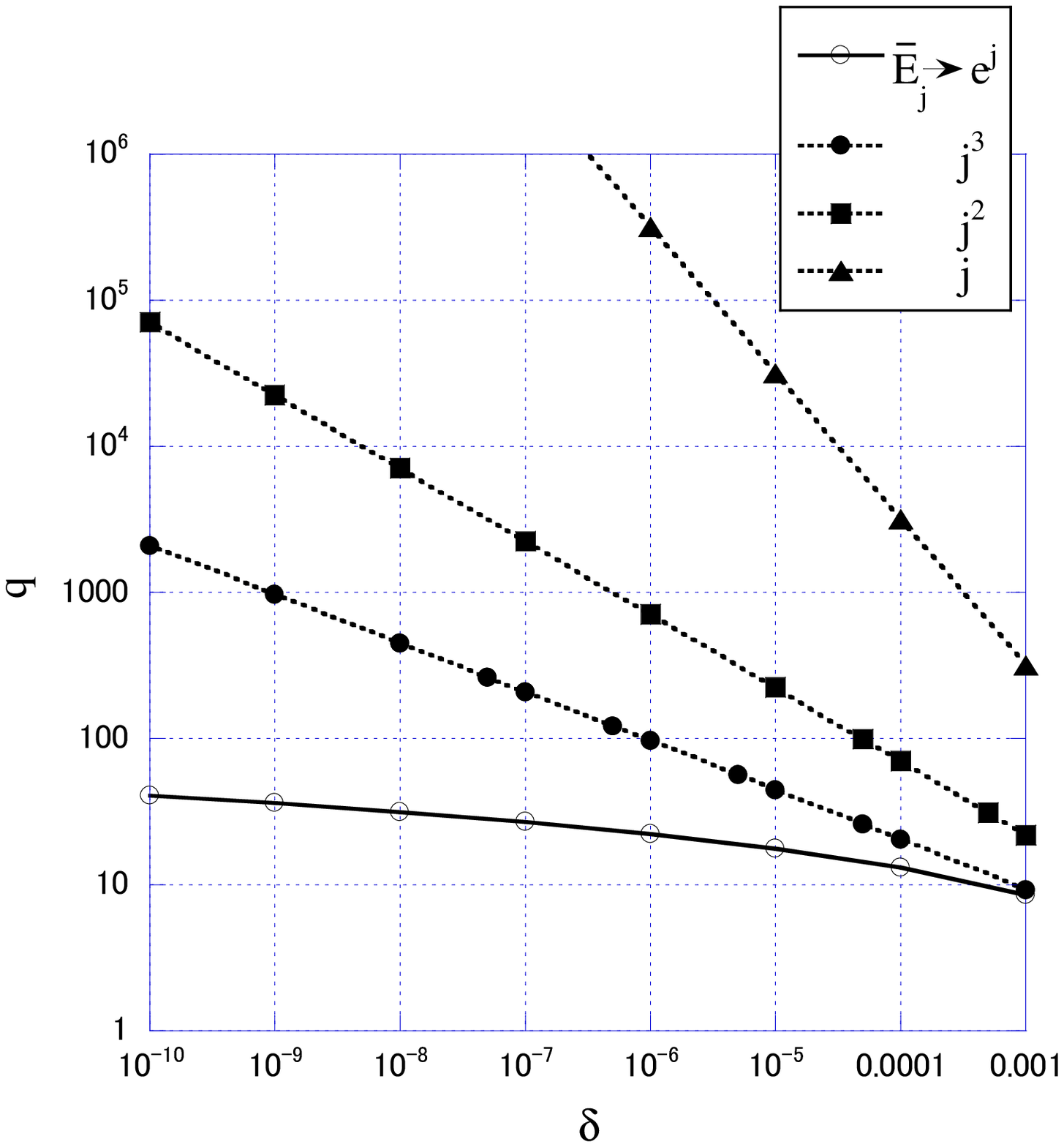,width=3in}
\psfig{file=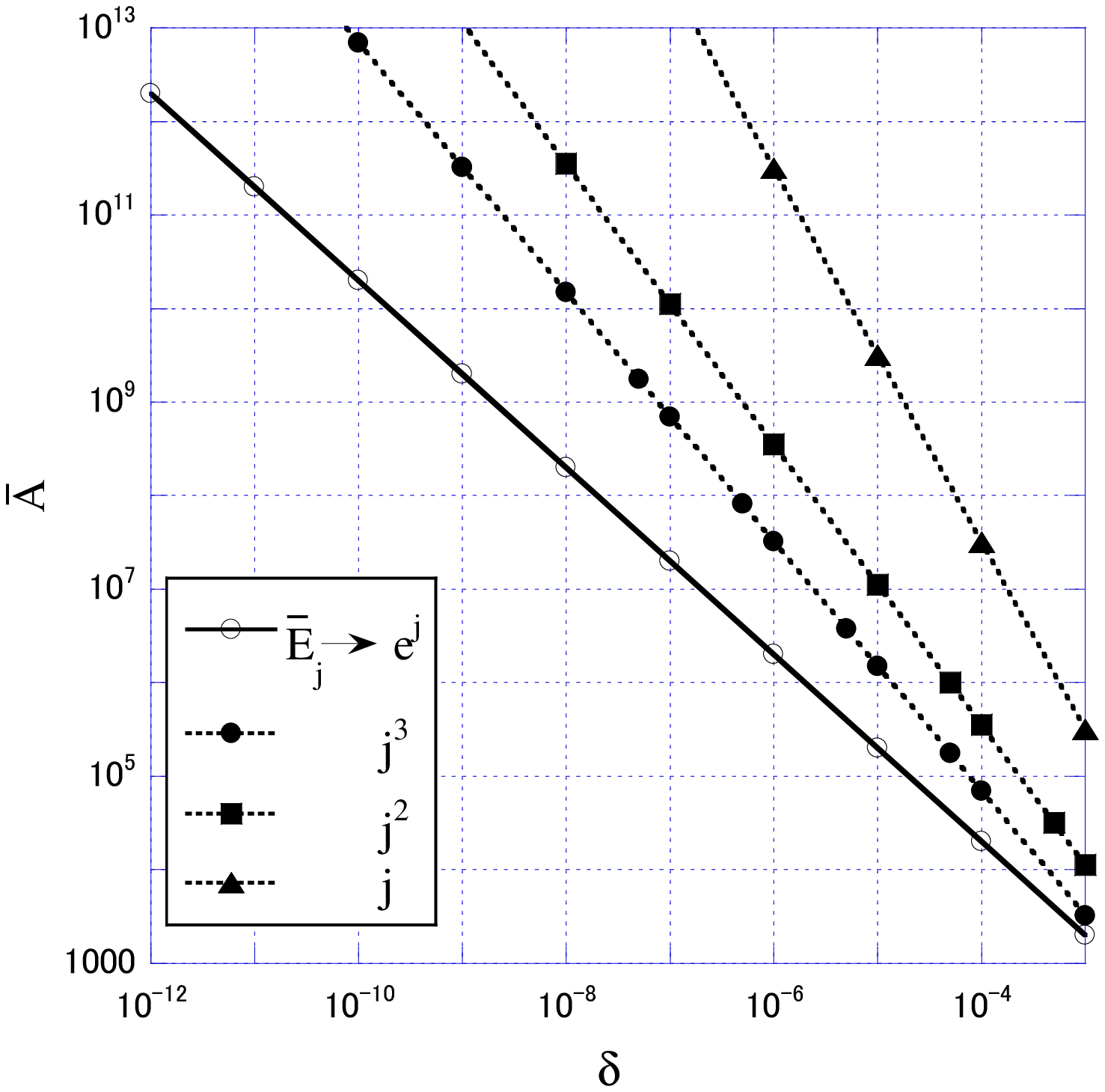,width=3in}
\caption{\label{delta-MB}
$\delta$-$q$ and $\bar{A}$ relations showing the consistency with (\ref{delta-q-relation}), 
(\ref{delta-A-relation}), (\ref{delta-q-relation2}), and (\ref{delta-A-relation2}).  }
\end{figure}

If we use the relations $(\Delta E)^{2}=-\partial_{\beta}E$, (\ref{quasilocal-1st}), and 
(\ref{delta-A-relation}), we obtain 
\begin{eqnarray}
\frac{\Delta E}{E}=\frac{\Delta A}{A}\propto \delta^{\frac{1}{2n}}\ .
\label{fluctuation}
\end{eqnarray}
The case of (\ref{delta-A-relation2}) is included in the limit $n\to \infty$. 
It is surprising that fluctuations of both energy and horizon area 
are summarized in this simple manner. 

In the above estimate, we used the asymptotic form in the limit $j\to\infty$. 
Thus, it is desirable to check consistency using a numerical calculation. 
For this purpose, we choose 
\begin{eqnarray}
\rho^{2}=0,\ j^{4},\ j^{6},\ e^{2j}, 
\label{function-form}
\end{eqnarray}
which correspond to the cases, 
\begin{eqnarray}
\bar{E}_{j}\to j,\ j^{2},\ j^{3},\ e^{j}, 
\label{energy-asymptotic}
\end{eqnarray}
in the limit $j\to\infty$, respectively. However, we stress that we use the exact 
expression (\ref{E-spin}) by substituting (\ref{function-form}). 
We show $\delta$-$q$, $\bar{A}$ relations in Figs.~\ref{delta-MB} 
which have complete consistency with 
(\ref{delta-q-relation}), (\ref{delta-A-relation}), (\ref{delta-q-relation2}), 
and (\ref{delta-A-relation2}). Especially, 
in all cases, $A\to \infty$ for $\delta\to 0$. 
So we confirmed that the holographic bound is saturated, i.e., 
$\delta_{h}\to 0$, in the semiclassical limit where the temperature 
should approach Unruh temperature $\beta \to \beta_{\rm U}$. 
This is a generalization of the result in \cite{gas}. 

\begin{figure}[htbp]
\psfig{file=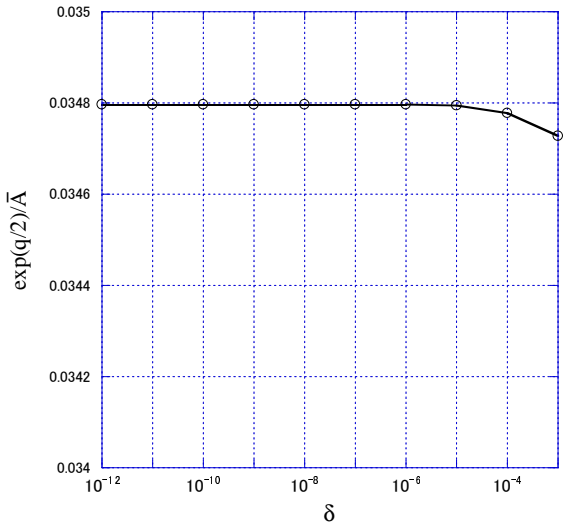,width=3.0in}
\caption{\label{delta-correct}
$\delta$-$\exp (q/2)/\bar{A}$ relation for $\rho^{2}=e^{2j}$. }
\end{figure}

Then, we should also notice the results $q\propto \ln \bar{A}$ for $\rho^{2}=e^{2j}$ derived by 
(\ref{delta-q-relation2}) and (\ref{delta-A-relation2}). To check its accuracy, 
we also show that $\exp (q/2)/\bar{A}$ 
is almost constant for $\rho^{2}=e^{2j}$ in Fig.~\ref{delta-correct}. 
The deviation from constant for large $\delta$ would be due to it from 
the asymptotic form. 
So, we obtain the $\log$ correction 
if we use the freedom $\rho^{2}$. This is also our new results 
obtained by considering the covariant area spectrum.

Finally, we comment on the case without the Gibbs factor. 
In this case, we have 
\begin{eqnarray}
Z_{\rm MB}=\sum q^{N}\ .
\label{without}
\end{eqnarray}
So, $q<1$ is required. However, it is impossible in the small $\delta$ as 
we see from (\ref{delta-q-relation}) and (\ref{delta-q-relation2}).

\section{Bose-Einstein statistics}
Here, we consider Bose-Einstein statistics as a candidate of the 
puncture statistics. First, we discuss the case $z=1$ as an extension 
of the case in Maxwell-Boltzmann statistics. In this case, the 
grand canonical partition function can be written as 
\begin{eqnarray}
Z_{\rm BE}(\beta)=\prod_{j}\left[ 1-\exp (-\delta\beta_{\rm U} E_{j}) 
\right]^{-1}\ .\label{grand-canonicalBE0}
\end{eqnarray}
So, we have 
\begin{eqnarray}
q:=\ln Z_{\rm BE}(\beta )=-\Sigma_{j}\ln \left[ 1-\exp (-\delta\beta_{\rm U} E_{j}) 
\right]\ .\label{lnBE0}
\end{eqnarray}
We can perform an analogous discussion in the previous section. 
For example, if we have $\bar{E}_{j}\to j^{n}$ $(n=1,m)$ in $j\to\infty$, 
we replace $f(x)$ by $g(x)=\ln \left[ 1-\exp (-2\pi x^{n}) 
\right]$ in (\ref{correction-key}). 
Then, the discussion below (\ref{correction-key}) holds, and we obtain 
(\ref{delta-q-relation}) and (\ref{delta-A-relation}). Similarly, 
for $\bar{E}_{j}\to e^j$, we obtain (\ref{delta-q-relation2}) 
and (\ref{delta-A-relation2}). 

The conclusions are that we have a holographic bound in the large area limit, and the 
correction term of the entropy-area law behaves same as the case in Maxwell-Boltzmann 
statistics qualitatively. The result for $n=1$ is consistent with \cite{gas} 
where the correction term is shown to be proportional to $\sqrt{A}$ both in 
Maxwell-Boltzmann statistics and in Bose-Einstein statistics. 
We have shown that these can be understood in an unified way including the cases in 
covariant area spectrum. 

Next, we discuss the case $z\neq 1$. 
The grand canonical partition function can be written as 
\begin{eqnarray}
Z_{\rm BE}(\beta ,\mu )=\prod_{j}\left[ 1-\exp (\beta\mu-\delta\beta_{\rm U} E_{j}) 
\right]^{-1}\ .\label{grand-canonicalBE}
\end{eqnarray}
So, we should require 
\begin{eqnarray}
\delta\beta_{\rm U} E_{j}-\beta\mu >0\ .\label{positiveBE}
\end{eqnarray}
Since we assumed that $E_{j}$ is monotonic 
function with $j$, we obtain
\begin{eqnarray}
{\rm if}\ \beta >\beta_{\rm U}(1-\delta_{h})\ ,\ \ {\rm then}\ \ 
\mu <\frac{\beta_{\rm U}\delta}{\beta} E_{1/2}\ ,
\nonumber\\
{\rm if}\ \beta <\beta_{\rm U}(1-\delta_{h})\ ,\ \ {\rm then}\ \ 
\mu =-\infty\ \ ,\nonumber  
\end{eqnarray}
as an extension of \cite{gas}. 
So, the high temperature region with $\beta <\beta_{\rm U}(1-\delta_{h})$ should be 
described by a Maxwell-Boltzmann statistics. We concentrate on the 
case with $\beta >\beta_{\rm U}(1-\delta_{h})$. 

We consider whether or not above discussion can be extendible for the case $z\neq 1$. 
We define 
\begin{eqnarray}
q:=\ln Z_{\rm BE}=-\Sigma_{j}\ln \left[ 1-\exp (\beta\mu -\delta\beta_{\rm U} E_{j}) 
\right]\ .\label{lnBE2}
\end{eqnarray}
The rhs of this equation includes two independent parameters $\delta_{\beta}$ 
and $\delta_{h}$. To avoid complication, we set $\delta_{h}=0$ below. 
Then, if we have $\bar{E}_{j}\to j^{n}$ in $j\to\infty$, 
we replace $g(x)$ by 
$g'(x)=\ln \left[ 1-\exp \left(-2\pi x^{n}+\beta\mu\right) \right]$ to 
perform an analogous discussion. 

However, in this case, $g'(x)$ does not necessarily satisfy {\bf Conditions} in 
the previous section. 
This depends on the ratio between $\left(2\pi x_{1}^{n}-\beta\mu\right)$ 
and $\epsilon$. In concrete, if $\left(2\pi x_{1}^{n}-\beta\mu\right)$ is 
small enough, $g'(x_{1}+\frac{\epsilon}{2})/g'(x_{1})$ can be much smaller than $1$. 
Of course, if we take $\epsilon\to 0$, we can obtain same conclusion as above. 
Below, we consider the case where {\bf Conditions} are violated. 

\begin{figure}[b]
\psfig{file=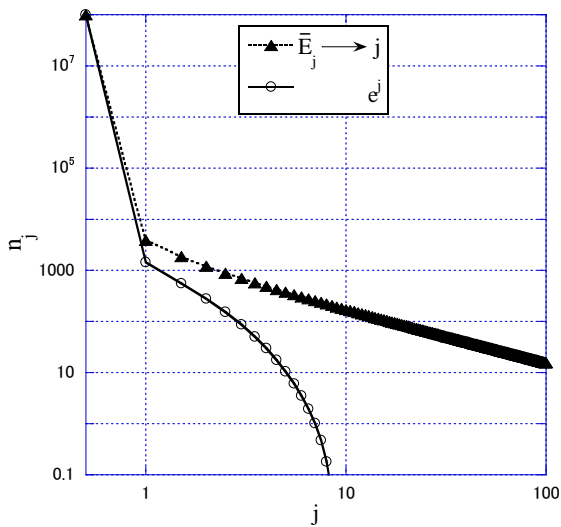,width=3.0in}
\caption{\label{number4j}
The relation between $j$ and its number density 
corresponding to $\bar{E}_{j}\to j$ and $e^{j}$ 
for $\delta =10^{-4}$, $\alpha =10^{-8}$ and $\delta_{h}=0$. }
\end{figure}

We can understand physical meaning of {\bf Conditions} by using the number of 
punctures $n_{j}$ for general $\bar{E}_{j}$. Here, $n_{j}$ is represented by 
\begin{eqnarray}
n_{j}=\left[ \exp (\delta \beta_{\rm U}E_{j}-\beta\mu ) -1
\right]^{-1}\ .\label{number}
\end{eqnarray}
We define
\begin{eqnarray}
\alpha :=\delta \beta_{\rm U}E_{1/2}-\beta\mu \ .\label{BE-parameter}
\end{eqnarray}
If $\alpha \ll 1$, the mean number of the grand state can be approximated as  
\begin{eqnarray}
n_{1/2}\simeq \frac{1}{\alpha}\ ,\label{grand-number}
\end{eqnarray}
which is quite large. Thus, $n_{1}/n_{1/2}\ll 1$ is possible, which corresponds to 
the case $g'(x_{1}+\frac{\epsilon}{2})/g'(x_{1})\ll 1$. 

Moreover, the total sum of the mean number $j>1/2$, $n_{\rm ex}:=\sum_{j=1}^{\infty}n_{j}$ 
can be much smaller than $n_{1/2}$ which 
corresponds to the Bose-Einstein condensate state defined 
as a state where the horizon is almost dominated by spin $1/2$ puncture.  
Since $n_{j}$ ($j\geq 1$) can satisfy {\bf Conditions}, 
$n_{\rm ex}$ can be estimated 
by following the analogous discussion as above. That is, 
if we have $\bar{E}_{j}\to  j^{n}$ or $e^{j}$ in $j\to\infty$, 
we can estimate that $n_{\rm ex}\propto \delta^{-1/n}$ or 
$ -\ln \delta $, respectively. 
So the criteria for the Bose-Einstein condensate are 
\begin{eqnarray}
{\rm if}\ \bar{E}_{j}\to  j^{n}\ ,\ \ {\rm then}\ \ 
\frac{1}{\delta^{1/n}} \ll \frac{1}{\alpha} \ ,
\label{BE1}\\
{\rm if}\ \bar{E}_{j}\to e^{j}\ ,\ \ {\rm then}\ \ 
-\ln\delta\ll \frac{1}{\alpha}\ \ .\label{BE2}  
\end{eqnarray}

We show the relation between $j$ and its number density 
corresponding to $\bar{E}_{j}\to j$ or $e^{j}$ 
for $\delta =10^{-4}$ and $\alpha =10^{-8}$ in Fig.~\ref{number4j}. 
Although both are the cases of the Bose-Einstein 
condensate, decays of $n_{j}$ make a contrast in these cases.  

\begin{figure}[b]
\psfig{file=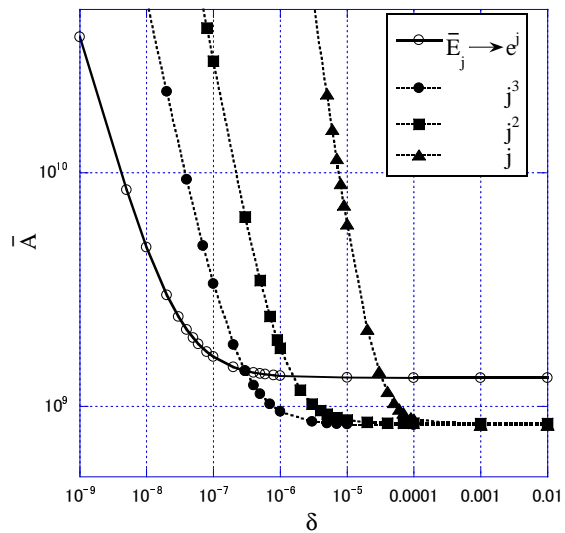,width=3.0in}
\caption{\label{delta-BE}
We show $A$ as a function of $\delta$ 
for the case $\alpha =10^{-8}$ and $\delta_{h}=0$. }
\end{figure}

We are interested in changes of physical quantities caused by 
the Bose-Einstein condensate. 
We show $\bar{A}$ as a function of $\delta$ 
for the case $\alpha =10^{-8}$ in Fig.~\ref{delta-BE}. 
Surprisingly, plateau appears for large $\delta$ while $\bar{A}$ increases as 
$\delta\to 0$ following 
(\ref{delta-A-relation}) or (\ref{delta-A-relation2}) for small $\delta$. 
If we use the criteria (\ref{BE1}) and (\ref{BE2}), the Bose-Einstein condensate 
occurs for all $\delta$ in this diagram. Then, 
how can we understand this plateau? 

We can discuss that $\bar{A}$ in the plateau 
corresponds to the case where $\bar{A}$ almost consists of the area spectrum 
$j=1/2$, $\bar{A}_{1/2}$. The reason is as follows. 
To estimate the area $\bar{A}_{\rm ex}$ consisting of the area spectrum $j>1/2$, 
we use 
\begin{eqnarray}
\frac{\bar{A}_{\rm ex}}{\beta}=E_{\rm ex}=n_{\rm ex}\mu 
-\partial_{\beta}\ln Z_{\rm BE,ex}\ . \label{general-A}
\end{eqnarray}
If $\bar{E}_{j}\to  j^{n}$ in $j\to\infty$, we have 
\begin{eqnarray}
n_{\rm ex}\propto \delta^{-1/n}\ ,\ \ 
\partial_{\beta}\ln Z_{\rm BE,ex}\propto \delta^{-(n+1)/n}\ , 
\end{eqnarray}
where we used (\ref{delta-q-relation}) and (\ref{delta-A-relation}). 
So, if $\delta$ is small enough, first term of rhs in (\ref{general-A}) 
can be negligible. Thus, we have 
\begin{eqnarray}
\bar{A}_{\rm ex}\propto \delta^{-(n+1)/n}\ . \label{A-ex}
\end{eqnarray}
Similarly, we can consider the case $\bar{E}_{j}\to  e^{j}$ in $j\to\infty$ 
and this case is included 
in the limit $n\to\infty$ in (\ref{A-ex}). 
So the condition for $\bar{A}_{\rm ex}\ll \bar{A}_{1/2}\propto \alpha^{-1}$ 
can be estimated as 
\begin{eqnarray}
\delta \gg \alpha^{n/(n+1)}\ . 
\label{criterion}
\end{eqnarray}
We can find that this is consistent with the results in Fig.~\ref{delta-BE}. 
This result is also newly revealed in this paper. 

If we consider what observables in black hole physics are, we may adopt the criterion 
(\ref{criterion}) as a condition for the Bose-Einstein condensate. 
When this condition is broken, $\bar{A}$ shows rapid grow as $\delta\to 0$. 
If we can discuss this phenomena as a phase transition from the 
quantum black hole to the classical black hole, it is very interesting. 

\section{Conclusion and discussion}
We have investigated the puncture statistics based on the covariant area spectrum. 
First, we have considered Maxwell-Boltzmann statistics with a Gibbs factor for punctures. 
If we assume the fugacity $z\neq 1$, we have reconfirmed the results in \cite{gas} that 
the correction term of the entropy-area law disappears for $z=e$. 
When we assume the fugacity $z=1$, 
we have established formulae which relate physical quantities such as horizon area to the parameter 
characterizing holographic degrees of freedom using asymptotic form of the area spectrum in the 
large spin limit. We have also performed numerical calculations and obtained consistency 
with these formulae. From these results, 
we have obtained that the holographic bound is satisfied in the large area limit 
which is the extension of the previous research. We have found that the correction term of 
the entropy-area law can be proportional to the logarithm of the horizon area 
as it has been pointed out in other researches. 

Second, we have also considered Bose-Einstein statistics and shown that above formulae are also 
useful in this case. 
By applying the formulae, we have understood intrinsic features of the Bose-Einstein condensate which correspond 
to the case when the horizon area almost consists of punctures in the ground state. 
We have shown that when this phenomena occurs, 
the area is approximately constant against the parameter $\delta$ characterizing the temperature. 
When this phenomena is broken, the area shows a rapid increase as $\delta\to 0$, which suggests 
the phase transition from quantum to classical area. 

What should we consider as a next step? 
Although we have assumed that $\rho$ is a dependent function of $j$, the validity 
should be checked by other method. For example, 
to reveal the property of $\rho$ in the covariant area spectrum and the puncture 
statistics, it is important to investigate the 
Hawking radiation as in \cite{spectroscopy} which is one of our future work. 
It is also interesting to discuss possibility of the phase transition using 
covariant area spectrum as in \cite{phase}. 
In a long span, we should also investigate a covariant volume spectrum, which would 
lead us to the covariant loop quantum cosmology. 
This must be the interesting arena in the next decade.

\section*{Acknowledgements}

We would like to thank Kei-ichi Maeda for continuous encouragement. 
We are thankful for financial support from the Nihon University.



\begin{thebibliography}{99}
\bibitem{Wheeler}
For reviews, see e.g. J.\ J.\ Halliwell, in \textit{Quantum Cosmology and Baby Universes}, 
edited by S.\ Coleman, J.\ B.\ Hartle, T.\ Piran, and 
S.\ Weinberg (World Scientific, Singapore, 1991);
C.\ Kiefer, \textit{Quantum Gravity} (Clarendon Press, Oxford, 2004);
D.\ H.\ Coule, Class.\ Quantum\ Grav.\ \textbf{22}, R125 (2005). 
\bibitem{complex}
A. Ashtekar, Phys. Rev. Lett. {\bf 57}, 2244 (1986); {\it ibid.}, 
Phys. Rev. D {\bf 36}, 1587 (1987). 
\bibitem{Jacobson}
T. Jacobson and L. Smolin, Nucl. Phys. B {\bf 299}, 295 (1988). 
\bibitem{Smolin}
C. Rovelli and L. Smolin, Phys. Rev. D {\bf 52}, 5743 (1995). 
\bibitem{Rovelli}
C. Rovelli and L. Smolin, Nucl. Phys. B {\bf 442}, 593 (1995); 
Erratum, {\it ibid.}, {\bf 456}, 753 (1995). 
\bibitem{Ash1}
A. Ashtekar and J. Lewandowski, Class. Quantum Grav. {\bf 14}, 
A55 (1997). 
\bibitem{Immirzi}
J. F. Barbero G., Phys. Rev. D {\bf 51}, 5507 (1995); 
G. Immirzi, Nucl. Phys. Proc. Suppl. B {\bf 57}, 65 (1997). 
\bibitem{Thiemann's method}
T. Thiemann, Phys. Lett. B {\bf 380}, 257 (1996). 
\bibitem{LQC}
M. Bojowald, Living. Rev. Rel. {\bf 8}, 11 (2005);
A. Ashtekar, T. Pawlowski, P. Singh, Phys. Rev. D {\bf 74}, 084003, (2006). 
\bibitem{Rovelli-entropy}
C. Rovelli, Phys. Rev. Lett. {\bf 77}, 3288 (1996). 
\bibitem{ABCK} 
A. Ashtekar, J. Baez, A. Corichi, and K. Krasnov, 
Phys. Rev. Lett. {\bf 80}, 904 (1998); 
A. Ashtekar, J. Baez, and K. Krasnov, Adv. Theor. Math. 
Phys. {\bf 4}, 1 (2000). 
\bibitem{isolated}
A. Ashtekar, A. Corichi, and K. Krasnov, Adv. Theor. Math. 
Phys. {\bf 3}, 419 (1999). 
\bibitem{Domagala}
M. Domagala and J. Lewandowski, Class. Quant. Grav. {\bf 21}, 5233 (2004); 
K. A. Meissner, {\it ibid.}, 5245 (2004). 
\bibitem{Alekseev}
A. Alekseev, A. P. Polychronakos, and M. Smedback, Phys. Lett. B 
{\bf 574}, 296 (2003); A. P. Polychronakos, Phys. Rev. D {\bf 69}, 
044010 (2004). 
\bibitem{Mitra}
A. Ghosh and P. Mitra, Phys. Lett. B {\bf 616}, 114 (2005); 
Phys. Rev. D {\bf 74}, 064026 (2006). 
\bibitem{Tamaki}
T. Tamaki and H. Nomura, Phys. Rev. D {\bf 72}, 107501 (2005); 
T. Tamaki, Class. Quant. Grav. {\bf 24}, 3837 (2007); 
T. Tanaka and T. Tamaki, Eur. Phys. J. C, {\bf 73}, 2314 (2013). 
\bibitem{Schiappa}
For review, see, e.g., J. Natario and R. Schiappa, 
Adv. Theor. Math. Phys. {\bf 8}, 1001 (2004). 
\bibitem{Dreyer}
O. Dreyer, Phys. Rev. Lett. {\bf 90}, 081301 (2003); S. Hod, Phys. Rev. Lett. {\bf 81}, 4293 (1998). 
\bibitem{Hod}
T. Tamaki and H. Nomura, Phys. Rev. D {\bf 70}, 044041 (2004); 
H. Nomura and T. Tamaki, Phys. Rev. D {\bf 71}, 124033 (2005). 
\bibitem{projection} J. F. Barbaro G. and E.J.S. Villasenor, 
Class. Quant. Grav. {\bf 26}, 035017 (2009). 
\bibitem{Sahlmann}
H. Sahlmann, Phys. Rev. D {\bf 76}, 104050, (2007);
H. Sahlmann, Class. Quant. Grav. {\bf 25}, 055004 (2008). 
\bibitem{Ansari}
M. H. Ansari, Nucl. Phys. B {\bf 783}, 179 (2007); {\it ibid.}, 
{\bf 795}, 635 (2008). 
\bibitem{Engle}
J. Engle, A. Perez, and K. Noui, Phys. Rev. Lett. {\bf 105}, 031302 (2010); 
J. Engle, K. Noui, A. Perez and D. Pranzetti, Phys. Rev. D {\bf 82}, 
044050 (2010); A. Perez and D. Pranzetti, Entropy {\bf 13}, 744 (2011).  
\bibitem{BTZ}
E. Frodden, M. Geiller, K. Noui, and A. Perez, JHEP {\bf 05}, 139 (2013). 
\bibitem{analytic}
E. Frodden, M. Geiller, K. Noui, and A. Perez, 
Euro. Phys. Lett. {\bf 107}, 10005 (2014); 
B. A. Jibril, A. Mouchet, and K. Noui, JHEP {\bf 06}, 145 (2015). 
\bibitem{temperature}
D. Pranzetti, Phys. Rev. D {\bf 89}, 104046 (2014). 
\bibitem{Samuel}
J. Samuel, Class. Quant. Grav. {\bf 17}, L141 (2000); 
Phys. Rev. D {\bf 63}, 068501 (2001). 
\bibitem{Choice}
S. Alexandrov, Phys. Rev. D {\bf 65}, 024011 (2001). 
\bibitem{areaCLQG}
S. Alexandrov and D. Vassilevich, Phys. Rev. D {\bf 64}, 044023 (2001). 
\bibitem{HilbertCLQG}
S. Alexandrov, Phys. Rev. D {\bf 66}, 024028 (2002). 
\bibitem{reviewCLQG}
For review, see, e.g., 
S. Alexandrov, M. Geiller, and K. Noui, SIGMA, 
{\bf 8}, 055 (2012); S. Alexandrov and P. Roche, 
Physics Reports, {\bf 506}, 41 (2011). 
\bibitem{countingCLQG}
S. Alexandrov, arXiv:gr-qc/0408033. 
\bibitem{first}
A. Ghosh and A. Perez, Phys. Rev. Lett. {\bf 107}, 241301 (2011). 
\bibitem{gas}
A. Ghosh, K. Noui, and A. Perez, Phys. Rev. D {\bf 89}, 084069 (2014); 
O. Asin, J. B. Achour, M. Geiller, K. Noui, and A. Perez, 
Phys. Rev. D {\bf 91}, 084005 (2015). 
\bibitem{quasilocal}
E. Frodden, A. Ghosh, and A. Perez, Phys. Rev. D {\bf 87}, 121503(R) (2013). 
\bibitem{Bombelli}
L. Bombelli, R. K. Koul, J. Lee, and R. D. Sorkin, Phys. Rev. D {\bf 34},
 373 (1986). 
\bibitem{Carlip}
S. Carlip, Class. Quant. Grav. {\bf 17}, 4175 (2000). 
\bibitem{Das}
S. Das, P. Majumdar, and R. K. Bhaduri, Class. Quant. Grav. {\bf 19}, 
2355 (2002). 
\bibitem{Sen}
A. Sen, J. High Energy Phys. {\bf 04} (2013) 156. 
\bibitem{spectroscopy}
A. Barrau, X. Cao, K. Noui, and A. Perez, 
Phys. Rev. D {\bf 92}, 124046 (2015). 
\bibitem{phase}
J. M\"akel\"a, Phys. Rev. D {\bf 93}, 084002 (2016). 
\end{thebibliography}
\end{document}